\documentclass[letterpaper, 10 pt, conference]{ieeeconf}  %

\IEEEoverridecommandlockouts                              %

\usepackage{graphicx} %
\usepackage{textcomp}

\usepackage[
    backend=biber,
    style=ieee,
    natbib=true,
    url=false, 
    doi=false,
    eprint=false
]{biblatex}
\renewcommand*{\bibfont}{\small}

\title{\LARGE \bf
Y-Net for Chest X-Ray Preprocessing: Simultaneous Classification of Geometry and Segmentation of Annotations
}

\author{John E. McManigle$^{1}$, Raquel R. Bartz$^{1}$, and Lawrence Carin$^{2}$%
\thanks{Work supported by the Duke University Department of Anesthesiology}%
\thanks{$^{1}$John E. McManigle and Raquel R. Bartz, Department of Anesthesiology,
        Duke University, DUMC 3094, Durham, NC 27710, USA
        {\small {\tt john.mcmanigle} and {\tt raquel.bartz@duke.edu}}}%
\thanks{$^{2}$Lawrence Carin, Department of Electrical and Computer Engineering, Duke University, Box 90291
        Durham, NC 27710, USA
        {\tt\small lcarin@duke.edu}}%
}

\begin{document}

\maketitle
\thispagestyle{empty}
\pagestyle{empty}

\begin{abstract}

Over the last decade, convolutional neural networks (CNNs) have emerged as the leading algorithms in image classification and segmentation.  Recent publication of large medical imaging databases have accelerated their use in the biomedical arena.  While training data for photograph classification benefits from aggressive geometric augmentation, medical diagnosis -- especially in chest radiographs -- depends more strongly on feature location.  Diagnosis classification results may be artificially enhanced by reliance on radiographic annotations.  This work introduces a general pre-processing step for chest x-ray input into machine learning algorithms.  A modified Y-Net architecture based on the VGG11 encoder is used to simultaneously learn geometric orientation (similarity transform parameters) of the chest and segmentation of radiographic annotations.  Chest x-rays were obtained from published databases.  The algorithm was trained with 1000 manually labeled images with augmentation. Results were evaluated by expert clinicians, with acceptable geometry in 95.8\% and annotation mask in 96.2\% ($n=500$), compared to 27.0\% and 34.9\% respectively in control images ($n=241$).  We hypothesize that this pre-processing step will improve robustness in future diagnostic algorithms.
\newline

\indent \textit{Clinical relevance}— This work demonstrates a universal pre-processing step for chest radiographs -- both normalizing geometry and masking radiographic annotations -- for use prior to further analysis.
\end{abstract}

\section{INTRODUCTION}
\subsection{Network Architecture}
Convolutional neural networks (CNNs) are recognized as the state-of-the-art algorithms for most forms of automated image analysis.  While the foundations of artificial neural networks were laid \citep{mccullochLogicalCalculusIdeas1943} in the 1940s, followed by forays into imaging applications \citep{rosenblattPerceptronPerceivingRecognizing1957} and development of backpropagation \citep{werbosBackpropagationTimeWhat1990,werbosRegressionNewTools1974} in the 1950s\,-\,80s, practical applications awaited hardware advances realized in the early 21st century.

CNNs are multi-layer networks that reduce overfitting and take advantage of structural information in images using convolutional layers, which apply perceptrons across overlapping, restricted input fields.  The power of this technique was famously demonstrated by the success of AlexNet \citep{krizhevskyImageNetClassificationDeep2017} in the 2012 ImageNet Large Scale Visual Recognition Challenge, a photograph classification task.  Deeper CNNs continued to dominate the challenge until it was discontinued in 2017, as algorithms outperformed humans.

Seeking to generalize CNNs to perform image segmentation, Shelhamer et al.\ proposed \citep{shelhamerFullyConvolutionalNetworks2017} fully convolutional networks (FCNs) in 2014.  In this architecture, the fully connected layers that terminate a classification CNN are replaced with a pixel-wise classification layer.  The FCN authors proposed pooling methods to upsample segmentation from the resolution of the final convolutional layer to that of the input image. This problem was further addressed by the development \citep{ronnebergerUNetConvolutionalNetworks2015} of U-Net by Ronneberger et al.\ in 2015.  In this architecture, the higher-resolution convolutional ``contracting path'' outputs are bridged and concatenated with upsampled ``expanding path'' outputs to take advantage of the resolution of the former and the accuracy of the latter.

Some image analysis problems require both whole-image outputs (classification) as well as localized outputs (i.\,e.\ segmentation).  In such cases, a classification architecture and its corresponding U-Net can be combined for the dual purpose into a ``Y-Net'' as proposed by Mehta et al.\ \citep{mehtaYNetJointSegmentation2018}  (Of note, the term ``Y-Net'' has also been used to describe other branching neural network architectures \citep{lanYNetHybridDeep2019,mohammedYNetDeepConvolutional2018}.)  Alternative strategies include Yang's multi-task DCNN \citep{yangNovelMultitaskDeep2017}, which uses a separate branch for each classification task, and He's Mask R-CNN \citep{heMaskRCNN2018}, which performs instance segmentation.

\begin{figure*}
	\centering
	\includegraphics[width=0.9\textwidth]{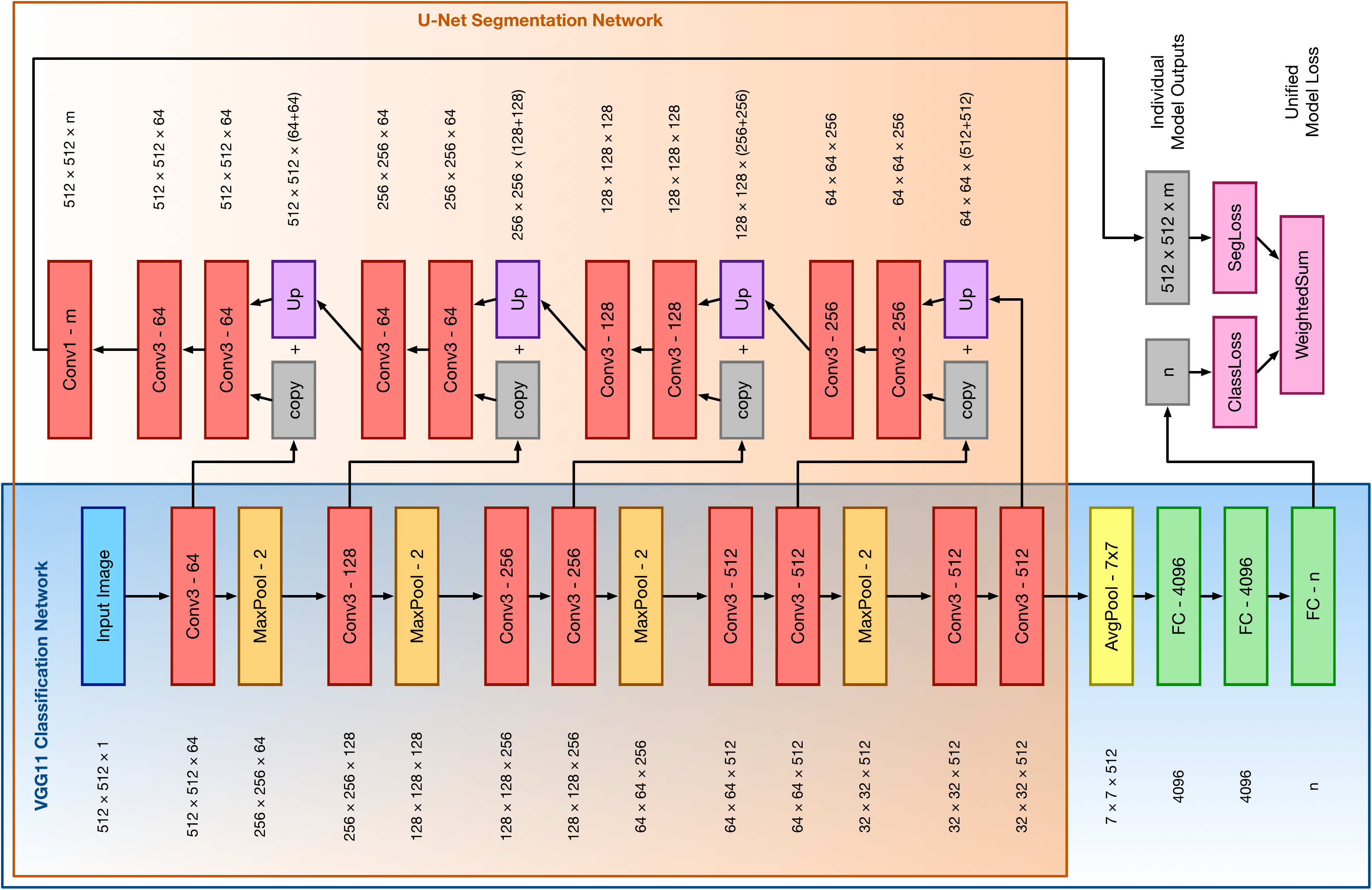}
	\caption{Our Y-Net architecture based on VGG11. Large blue box: classification network; large orange box: U-Net segmentation network; $n$:~classification channels; $m$:~segmentation channels; ``Conv3-$q$'': $3 \times 3$ convolution with padding, $q$-d output; ``MaxPool - 2'': $2 \times 2$ max pooling; ``AvgPool - $7 \times 7$'': $2 \times 2$ max pooling followed by average pooling with $7 \times 7$ output; ``FC'': fully connected; ``Up'': $2\times$ upsampling.  Dimensions represent the output size at each layer assuming a $512 \times 512 \times 1$ grayscale input image.}
	\label{fig:architecture}
\end{figure*}

\subsection{Chest X-Ray Classification}
After the huge successes in photograph classification, common wisdom was that classification (i.\,e.\ diagnosis) of medical images was soon to follow.  This progress has been hindered by a sparsity of available training data, commonly attributed to the difficulty of publishing datasets given patient privacy regulations as well as the expertise and/or expense involved in radiologic diagnosis.

Breakthroughs on the training data front occurred over the past two years, with release of large databases of chest radiographs with diagnostic information mined from radiologists' text reports.  ChestX-ray8 \citep{wangChestXRay8HospitalscaleChest2017} released by the National Institutes of Health, CheXpert \citep{irvinCheXpertLargeChest2019} released by Stanford University, and MIMIC-CXR \citep{johnsonMIMICCXRLargePublicly2019} released by Beth Israel Hospital together contain over 700,000 radiographs of over 150,000 individual patients.

Compared to the classification of photographs, classification of radiographs -- especially chest radiographs -- is differentiated by the importance of feature location within an image.  While a dog is a dog whether it is on the left or right side of a photograph, radiographic density could represent pericardial effusion or hemothorax depending on its location.  This prompts the hypothesis that the aggressive augmentation techniques using geometrically transformed training data that is so effective in photograph classification \citep{mikolajczykDataAugmentationImproving2018} may be less effective than methods that preserve anatomical position when considering radiographs.

One final pitfall of large-scale machine learning on radiographic data is the common use of annotations -- ``burned in'' to film or added electronically to the image.  These annotations include left/right side markers, notations of patient position or radiography technique, and patient identifiers.  Just as models trained to differentiate photographs of husky dogs from those of wolves famously relied on background information in the image \citep{ribeiroWhyShouldTrust2016}, models trained on radiographs -- especially multi-site datasets -- use annotations and other image features \citep{zechVariableGeneralizationPerformance2018} to deduce the hospital, ward, and even individual imaging machine in use, which provides brittle patterns then learned by the model \citep{rechtCIFAR10ClassifiersGeneralize2018}.

The object of the following work is to train a simple Y-Net to normalize the anatomical position of chest radiographs and segment radiographic annotations.  We hypothesize that such a pre-processing step will decrease overfitting and improve performance of a variety of future analyses of these films.

\section{METHODS}

\subsection{Experimental Data}
A combined database was created from the published ChestX-ray8, CheXpert, and MIMIC-CXR databases containing 707,626 images.  Of these, 553,344 are labeled as frontal chest films.  In consideration of future work, we reserved 52,102 images for future testing use, and selected all data (training, validation, and testing) in this experiment from the 501,242 remaining frontal chest films.

One thousand images were randomly selected to serve as training data.  These were labeled with four points (``top'': thoracic inlet, ``bottom'': center of chest at level of diaphragm, ``left'' and ``right'': lateral ribcage at the mid-thoracic level) by a clinical expert.  These marks were used to compute a four-parameter similarity geometry: the center point was defined as the midpoint between ``top'' and ``bottom'' points; ``rotation'' was the angle between the chest's vertical axis (line between ``top'' and ``bottom'' points) and the image vertical axis; and chest ``size'' was the larger of the distance between ``top'' and ``bottom'' points, and the distance between ``left'' and ``right'' points orthogonal to the chest's vertical axis.  These similarity parameters were ground truth for the classification limb of the network.

Additionally, bounding boxes were drawn around areas identified as ``radiographic annotation'' in the image, including fiduciary markers and overlaid text identifying.  These boxes were converted into ground truth masks for the segmentation limb of the network.

Of the 1,000 training images, 87 were excluded because they did not have sufficient clinical information to confidently label points.  The 913 labeled images were aggressively augmented with random similarity transforms: rotation from -90\textdegree{} to 90\textdegree{}, scale from 75\% to 125\% original size, and translation up to 25\% of the smaller image dimension.  Geometric parameters and annotation mask were transformed appropriately.  This resulted in a total of 600,000 images in the augmented dataset.

\subsection{Model Design and Training}

In order to simplify design as much as practical for this well-defined task, a simple Y-Net was built using a VGG11 backend.  The Y-Net is composed of the U-Net encoding and decoding networks, with the VGG11 fully connected classification branch spliced onto the encoding network (figure~\ref{fig:architecture}).  For simplicity, additional skip connections added by Mehta et al.\ were not used, so the classification and U-Net paths are identical to their vanilla VGG11 counterparts.

The network was implemented\footnote{Code is available at https://github.com/mcmanigle/GeoMask-Y-Net} in PyTorch \citep{paszkePyTorchImperativeStyle2019}, with reference to Alexandre's U-Net implementation \citep{alexandrePytorchUNet2019}.  The encoding network was pre-initialized with weights from the VGG11 classifier trained on ImageNet.  The 600,000-image augmented dataset was split into training (95\%, 570,000 images) and validation (5\%, 30,000 images) sets.  Mean squared error and soft margin loss \citep{wangEnsembleSoftmarginSoftmax2018} were chosen as the loss functions for the classification and segmentation branches of the network, respectively.  The individual loss functions were normed for a scale of 0 to 1, and total loss for the system was an equally weighted average of the two.  Stochastic gradient descent was used with learning rate $10^{-3}$ and momentum 0.9.  Learning rate was decreased by a factor of 0.2 every 20 epochs.

\begin{figure}
	\centering
	\includegraphics[width=0.8\columnwidth]{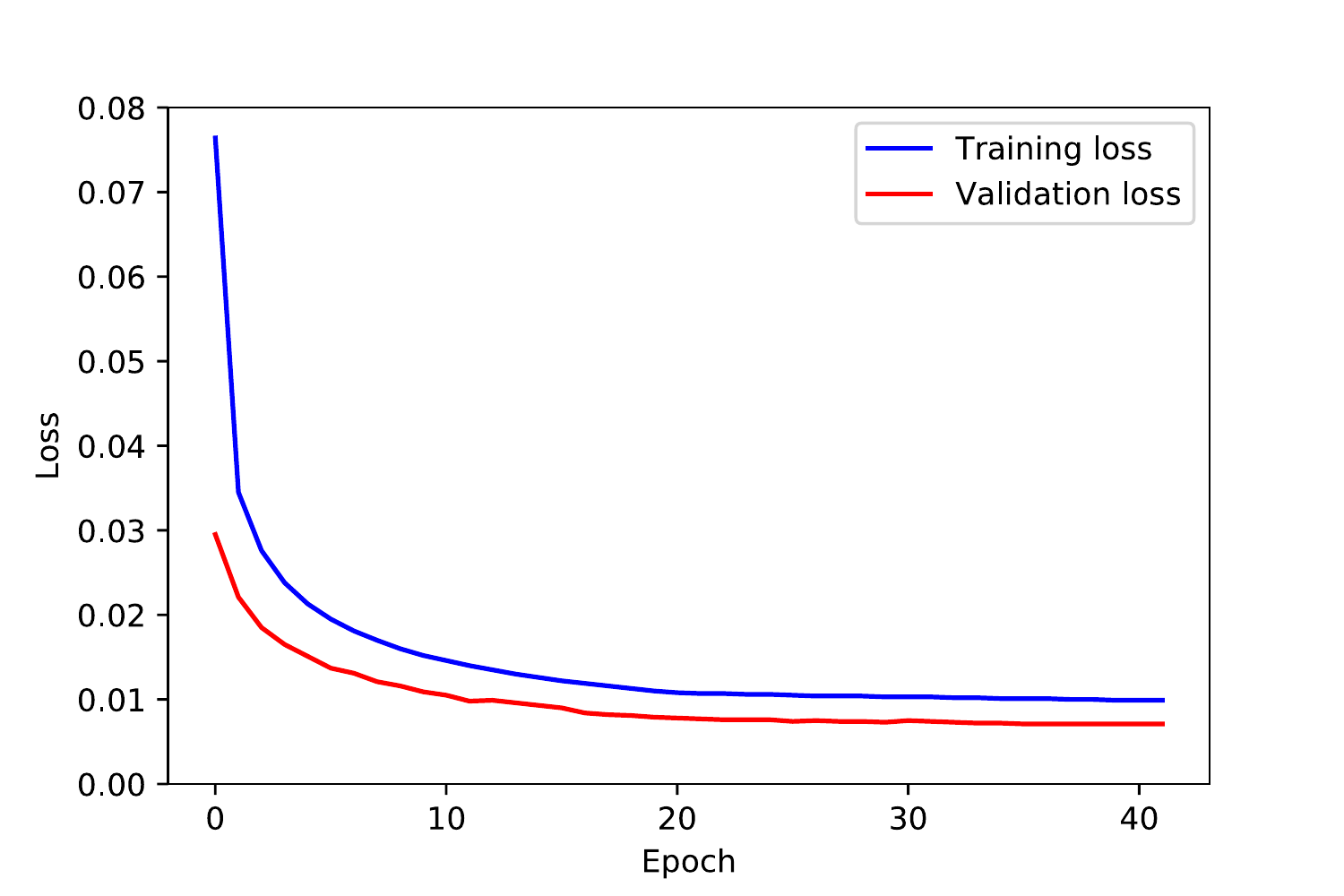}
	\caption{Training loss.}
	\label{fig:loss}
\end{figure}

The model was trained on an Amazon Web Services EC2 p3 instance (Amazon Web Services, Inc., Seattle, WA) with eight NVIDIA V100 Tensor Core GPUs (NVIDIA Corporation, Santa Clara, CA) running Ubuntu Linux (Canonical Group Ltd., London, UK).  Forty-two epochs were run over 76 hours (Figure~\ref{fig:loss}).

\section{RESULTS}

\begin{figure}
	\centering
	\includegraphics[width=0.82\columnwidth]{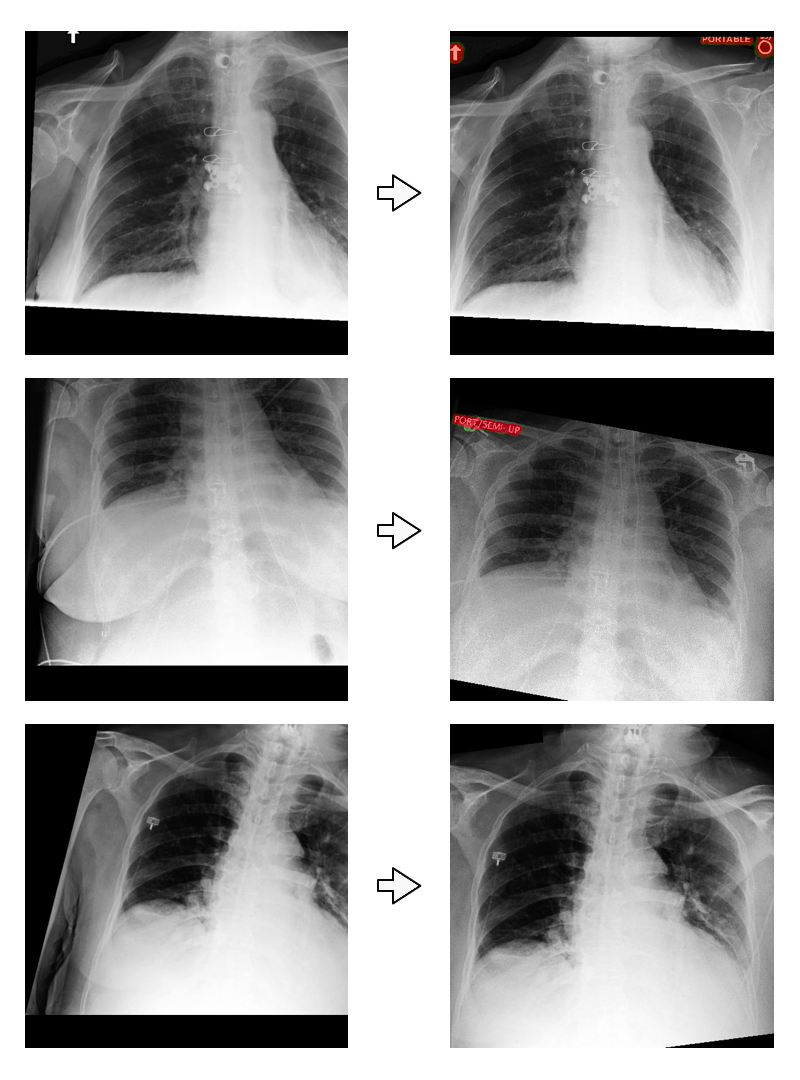}
	\caption{Example input radiographs (left) with output alignment (right) and annotation masks (red).}
	\label{fig:examples}
\end{figure}

After the model was trained, it was run on the full cohort of 501,242 frontal chest films (Figure~\ref{fig:examples}) for future evaluation purposes.  No further filtering (e.g.\ segmentation size threshold of annotation objects) was performed on the model output.

 For preliminary evaluation, images were randomly selected from images that were not included in the training set for evaluation by a group of trained physicians with specialized training in radiology, pulmonology, or critical care.  Evaluation proceeded by overlaying geometrically aligned images on an annotated template image to assist in grading of similarity transform parameters and annotation mask.  Evaluators were also presented with unmodified images from the published databases, center cropped to square and scaled to full image width, as controls.  Evaluation proceeded in a blinded manner.

\begin{figure}
	\centering
	\includegraphics[width=0.8\columnwidth]{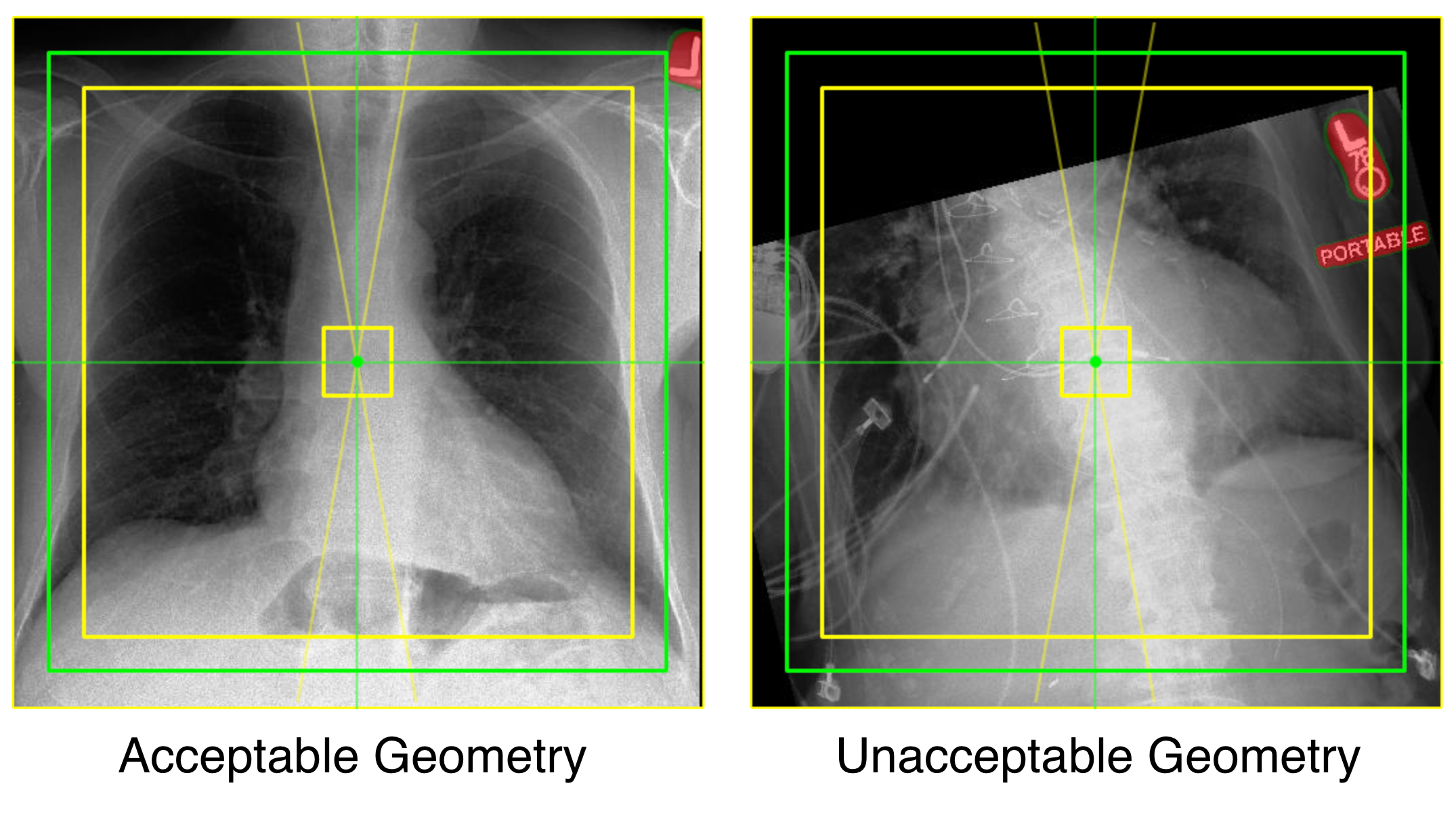}
	\caption{Example of acceptable and unacceptable images based on similarity geometry, shown with the gridline guides used by clinical evaluators.}
	\label{fig:geo}
\end{figure}

Images were scored as ``acceptable'' or ``unacceptable'' separately in both geometry and annotation mask (Figure~\ref{fig:geo}).  Images were aligned using a similarity transform to place the predicted center point at the center of the image, isotropically scale the radiograph such that the predicted larger dimension of the chest is equal to 90\% of the image width, and rotate the radiograph so that the vertical axis of the chest is upright.  After alignment, an image ``acceptable'' geometry meets all of the following criteria:
\begin{enumerate}
\item Center of the chest in the central 10\% of image.
\item Chest size (larger dimension) 80\%\,-\,100\% of image size, with no diagnostic area outside image.
\item Vertical axis of the chest within 10\textdegree{} of vertical.
\end{enumerate}
For images where patient position or image quality precluded meaningful applications of these rules, clinicians were asked to grade the image ``acceptable'' if its position was, in their opinion, ideal for interpretation, and ``unacceptable'' if a more appropriate position could be obtained.

Clinicians evaluated 241 control images and 500 experimental images. Geometry was acceptable in 65 (27.0\%) of the control images and 479 (95.8\%) of the experimental images.  Annotation mask was acceptable in 84 (34.9\%) of the control images -- representing images that did not have visible annotations -- and 481 (96.2\%) of the experimental images. ($p < 10^{-5}$ by Pearson's $\chi^2$.) Ongoing evaluation will characterize changes in statistical image information across the entire dataset.

\section{DISCUSSION}

This work serves as a proof of concept that a Y-Net can be trained to normalize geometry via a similarity transform and segment radiographic annotations.  Given the perils of future diagnostic models fitting to site in the current environment of few large published radiographic databases, we anticipate that decreasing non-clinical image cues -- including image orientation and radiographic annotations and markers -- could improve performance and increase reliance of models on clinically relevant data.  Additionally, orientation may allow models to use image location of features, with the tradeoff of decreasing the ability to perform aggressive image augmentation through geometric transforms.

\printbibliography

\end{document}